\documentclass[12pt]{aastex}
\usepackage{graphicx}
\usepackage{epsfig}
\usepackage{natbib}
\newcommand{\De}{\ensuremath{{\rm D}}}
\newcommand{\D}{\ensuremath{\null^2{\rm H}}}

\newcommand{\Het}{\ensuremath{\null^3{\rm He}}}
\newcommand{\He}{\ensuremath{\null^4{\rm He}}}

\newcommand{\Li}{\ensuremath{\null^7{\rm Li}}}

\begin{document}

\title{{\bf Helium and Deuterium Abundances as a Test for
the Time Variation of the Baryonic Density, Fine Structure
Constant and the Higgs Vacuum Expectation Value}}

\author{\bf \vspace{0.3cm} {
N. Chamoun \altaffilmark{1,2,a}, S. J. Landau
\altaffilmark{3,4,b}, M. E. Mosquera \altaffilmark{1,3,c}, Hector
Vucetich \altaffilmark{1,3,d}} {\vspace{0.3cm}}
$^1$\emph{Departamento de F{\'\i}sica, Universidad Nacional de La
Plata, cc67, 1900 La Plata, Argentina,}
\\
$^2$ \emph{Physics Department, HIAST, P.O.Box 31983, Damascus,
Syria,}
\\
$^3$\emph{Facultad de Ciencias Astron\'{o}micas y
  Geof\'{\i}sicas, Universidad Nacional de La Plata, Paseo del
Bosque, cp 1900 La Plata, Argentina,}\\$^4$\emph{Departamento de
F{\'\i}sica, FCEyN, Universidad de
  Buenos Aires, Ciudad Universitaria - Pab. 1, 1428 Buenos Aires,
  Argentina}\\}

\altaffiltext{a}{ nidal@fisica.unlp.edu.ar} \altaffiltext{b}{member of  the Carrera del Investigador Cient\'{\i}fico y Tecnol\'ogico, CONICET;
slandau@df.uba.ar} \altaffiltext{c}{fellow of CONICET; mmosquera@fcaglp.unlp.edu.ar}
\altaffiltext{d}{ vucetich@fcaglp.unlp.edu.ar}

\begin{abstract}

We use the semi-analytic method of \citet{Esma91} to calculate the
abundances of Helium and Deuterium produced during Big Bang 
nucleosynthesis assuming the fine structure constant and the Higgs
vacuum expectation value may vary in time. We analyze the
dependence on the fundamental constants of the nucleon mass,
nuclear binding energies and cross sections involved in the
calculation of the abundances. Unlike previous works, we do not
assume the chiral limit of QCD. Rather, we take into account the
quark masses and consider the one-pion exchange potential, within
 perturbation theory, for the proton-neutron scattering.
However, we do not consider the time variation of the strong
interactions scale but attribute the changes in the quark masses
to the temporal variation of the Higgs vacuum expectation value.
Using the observational data of the helium and deuterium, we put
constraints on the variation of the fundamental constants between
the time of nucleosynthesis and the present time.
%
\end{abstract}

\maketitle

\section{Introduction}
\label{sec:Intro} Big Bang Nucleosynthesis (BBN) offers the
deepest reliable probe of the early universe. Predictions of the
abundances of the light elements D,$^3$He, $^4$He and $^7$Li
synthesized at the end of the `first three minutes' are in good
overall agreement with the primordial abundances inferred from
observational data, which validates the Standard Big Bang
Nucleosynthesis (SBBN). BBN also provides powerful constraints on
possible deviations from the standard cosmology
and on new theories on physics beyond the Standard Model (SM)
\citep{sarkar96}.
Among these theories, there are those in which some of the
dimensionless ratios of fundamental constants do vary in time like
string-derived field theories
\citep{Wu86,Maeda88,Barr88,DP94,DPV2002a,DPV2002b}, related
brane-world theories
\citep{Youm2001a,Youm2001b,branes03a,branes03b}, {and} (related or
not) Kaluza-Klein theories
\citep{Kaluza,Klein,Weinberg83,GT85,OW97}. On the other hand,
recent astronomical data
\citep{Webb99,Webb01,Murphy01a,Murphy01b,Murphy03b} suggest a
possible variation of the fine structure constant
$\alpha=e^2/\hbar c$ at the $10^{-5}$ level over a time-scale of
10 billion years.  However, other recent independent analysis of
similar data \citep{MVB04,QRL04,Bahcall04,Srianand04,grupe05,chand06}
found no variation. On the other hand, measurements of molecular
hydrogen \citep{Ivanchik02, Ivanchik03,Ivanchik05} reported a
variation of the proton to electron mass $\mu = \frac{m_p}{m_e}$.
 This fact motivated
more general discussions of possible variations of other
constants.
 \citet{CF02} and \citet{Langacker02} have studied the implication of gauge
unification for the time variation of $\alpha$
while  \citet{Olive02} explored a super-symmetric version of the dynamical
Bekenstein model \citep{Bekenstein82}  in order to
produce a large change in $\alpha$ in the redshift range
$z=0.5-3.5$ and still be consistent with the constraints on
$\Delta \alpha / \alpha$ from the results of high precision limits
on the violation of equivalence principle by a fifth force.

On the other hand, there are many non-SBBN models which introduce new free
parameters in addition to the baryon density parameter, or equivalently
the baryon asymmetry $\eta_B \equiv \frac{n_B - n_{\bar{B}}}{n_\gamma}  =
2.74 \times 10^{-8}  \Omega_B h^2$. Most known of these models are those which
assume either a non-standard contribution to the total density, or a
lepton asymmetry. The first possibility affects the expansion rate of the
universe $S \equiv \frac{H'}{H}=\sqrt{\frac{\rho '}{\rho}}$ and can be
restated in terms of `equivalent' number of extra neutrinos $\Delta N_\nu =
N_\nu - 3$. Simple analytic fits
to BBN and  the cosmic
microwave background radiation (CMBR) data provide the following bound:  $0.85 < S < 1.15$  \citep{Barger03,Barger03b,Steigman05,Steigman06}. As
regards the lepton asymmetry, observational data do not imply that is
should be connected to the `tiny' baryon asymmetry $\eta_B$, and it could
be large enough to perturbe SBBN predictions .
Moreover, a small asymmetry between electron type
neutrinos and antineutrinos can have a significant impact on BBN since the
$\nu_e$ affect the interconversion of neutrons to protons changing the
equilibrium neutron-to-proton ratio from $(n/p)^0_{eq}=e^{-\frac{\Delta
m}{T}}$ to $(n/p)_{eq}=(n/p)^0_{eq} e^{-\xi_e}$. In consequence, the $\He$ abundance changes. In contrast the $\De$ abundance is  insensitive to $\xi_e \neq 0$. Consistent with the BBN and CMBR data, values of $\xi_e$ in the range
$-0.1 < \xi_e < 0.3$ are permitted \citep{Barger03,Steigman05,Steigman06}.
In our analysis, however, we shall not consider these non-SBBN scenarios,
but attribute any non-SBBN issue to time-variation of fundamental
constants.

The density of baryonic matter $\Omega_B h^2$
can be estimated using the WMAP data from the CMBR \citep{wmapest,wmap3}. From
the observed WMAP baryon density, the predicted abundances are
highly consistent with the observed $\De$ but not with $\He$ and
$\Li$. However, any change in the value of the fundamental
constants would work its way into the value of the abundances of
the various light elements and the question we address is whether
or not existing observations of the primordial abundances suggest
any change in the values of the fundamental constants at the time
of BBN.


BBN is sensitive to a number of fundamental dimensionless
parameters including the fine structure constant $\alpha$ ,
$\Lambda_{QCD}/M_{Plank}$ and  $m^q/\Lambda_{QCD}$ where $m^q$ is
the quark mass and $\Lambda_{QCD}$ is the strong scale determined
by the position of the pole in the perturbative QCD running
coupling constant. Several authors have studied the dependencies
of the BBN parameters on the fundamental constants. The dependence
of the primordial abundances on the fine structure constant has
been evaluated by \citet{Iguri99} and improved by \citet{Nollet}.
\citet{YS03} analyze the effects of the time variation of the
higgs vacuum expectation value $ < v >$ on BBN and the cosmic
microwave background radiation (CMBR). On the other hand,
\citet{mueller04} calculate the abundances as function of the
Planck Mass $M_P$, $\alpha$, $ < v
>$, electron mass $m_e$, nucleon decay time, deuterium binding
energy ($\epsilon_D$) and neutron-proton mass difference ($\Delta m=
m_n - m_p$). Moreover, they study the dependence of the last three
quantities as functions of the fundamental couplings and masses.
\citet{KM03} study the dependence of the primordial abundances on
$\epsilon_D$ and $\Delta m$. These papers \citep{YS03,mueller04,KM03} use the results of chiral perturbation theory  \citep{BS03} to adress the dependence of the deuterium binding energy with the higgs vacuum expectations value and $\Lambda_{QCD}$.  However, the dependence of the deuterium
binding energy $\epsilon_D$ on $<v>$ was estimated from an
approximated linear dependence of $\epsilon_D$ on the pion mass
$m_{\pi}$, while the exact limits on the relative change of $<v>$
would depend on the details of such dependence.  On the other hand,
\citet{CO95} and \citet{Ichi02} study the effects of variation of
fundamental constants on BBN in the context of a dilaton
superstring model. Finally, limits on cosmological variations of
$\alpha$, $\Lambda_{QCD}$ and $m^q$  from optical quasar
absorption spectra, laboratory atomic clocks and from BBN have
been established by \citet{Flambaum02,Flambaum04b}. For computing
the deuterium binding energy ($\epsilon_D \approx 2.225 \rm MeV$)
they apply quantum mechanics perturbation theory. This factor is
very significant in influencing the reaction rate of
$p+n\rightarrow d+\gamma$ which is the first and most crucial step
in BBN.

The BBN abundances can be computed using numerical
\citep{Wagoner73,Kawano92} and analytical
\citep{Esma91,Mukhanov03} methods. In a previous work
\citep{LMV06}, we used the semi-analytic method of \citet{Esma91}
to calculate the abundances of the light elements produced during
BBN assuming that the gauge coupling constants may vary in time.
We considered the chiral limit of QCD when analyzing the nucleon
masses, binding energies and the cross sections. Deviations
between standard cosmology calculations and observational data
could be interpreted as resulting from variations in $G_F$ the
Fermi constant, $\alpha$ the fine structure constant and
$\Lambda_{QCD}$ the strong interactions scale. The semi-analytical
method allows us to obtain semi-analytic dependencies of the
primordial abundances on the fundamental constants, which
otherwise must be computed using numerical codes.

On the other hand, in the standard model, a variation of the Fermi constant
implies a variation of the vacuum expectation value (vev) of the
Higgs field \citep{dixit88}:

\begin{equation}
G_F=\frac{\alpha_{weak}(M_W)}{\sqrt{2}M_W^2}=\frac{1}{\sqrt{2}
<v>^2}
\end{equation}
Here $M_W$ is the mass of the W-boson, and $<v> \approx 250
\rm{GeV}$ is the vev of the Higgs field. Within the QCD chiral
limit, the quark masses, which are also proportional to the Higgs
vev through the relation $m^q = Y_{Yukawa} <v>$, are neglected.
Therefore, the logical and consistent step to follow is to go
beyond the chiral limit and take the variation of $<v>$ as
affecting the Fermi constant as well as the quark masses.
We analyze the nucleon masses, the
nuclear binding energies and the cross sections dependence within
quantum mechanics perturbation theory. The objective of this paper is to study such variations as model independent as possible. Therefore,
we consider the one-pion exchange potential as the perturbation on
the p-n scattering responsible for the formation of the deuterium.
This perturbation potential varies in time if the pion mass
changes in time which leads to a time variation of the deuterium
binding energy. The pion mass also depends on the Higgs vev
through the Gell-Mann-Oakes-Renner relation: $m_{\pi}^2 = m^q
\frac{|<\bar{q}q>|}{f^2_\pi} \simeq m^q \Lambda_{QCD}$, where
$<\bar{q}q>$ is the quark condensate and $f_\pi$ is the pion decay
constant.
In order to determine the dependence of the deuterium binding
energy on the fundamental constants, we  use the square well model
to approximate the attractive strong interaction potential of the
deuterium and fit current scattering data to get estimates for the
depth and width of the well. On the other hand, we will not
discuss the effect of $\Lambda_{QCD}$ variation on the
QCD-determined quantities such as the quark condensate or the
width and depth of the square well. The reason for this is that we
lack a complete theory for these quantities, and, especially,
because of the absence of p-n scattering data in the far past. For
the same reason, we will not consider changes in the Yukawa
couplings either. On the other hand, the effect of  of $\Lambda_{QCD}$ variation on the abundances of the light elements was analyzed in a previous work \citep{LMV06}.
Thus, we will limit ourselves in this paper to
study the dependence on $\alpha$ and $<v>$ for the physical
quantities, such as binding energies, nucleon masses and cross
sections involved in the BBN calculations.  Our treatment of the
deuterium binding energy is similar to the one performed by of
\citet{Flambaum02}. However, there are some technical differences
in the wave function normalization which we describe in section
\ref{dbe}. Furthermore, we go one step further in calculating the
effects of such variations in the fundamental constants on the
primordial abundances of $D$ and $^4He$. This is one of the
advantages of using the semi-analytical method and it also allows
us to compare with observational data in order to put bounds on
the variation of $\alpha$ and $<v>$. 

On the other hand, the
concordance between the WMAP estimates and SBBN  has been investigated by many authors
\citep{cyburt03,romano03, cuoco03,cyburt04,coc04,Vangioni04}. From the
WMAP baryon density , the predicted abundances are highly consistent
with the observed $\De$ but not with $\He$ and $\Li$. They are
produced more than observed. Such discrepancy is usually ascribed to
non reported systematic errors in the observations of $\He$ and
$\Li$. Indeed, more realistic determinations of the $\He$ uncertainty
implies a baryon density in line with the WMAP estimate
\citep{cyburt04,Olive04}. On the other hand, \citet{richard05} have
pointed out that a better understanding of turbulent transport in the
radiative zones of the stars is needed for a better determination of
the $\Li$ abundance. In our previous work, we obtained results consistent with variation of fundamental constants when considering all data. However, discarding the $\Li$ data we obtained no variation. Therefore, we suspect that the possible non reported
systematic uncertainties are ``hidden'' within a setup involving variation of the
fundamental constants. Thus, until better estimations of the systematic errors of $\Li$ are reported,  we will only consider the $\De$ and $\He$ data.

Even though the WMAP estimate of the baryon density is the most
accurate one, it is still affected by degeneracies with other
cosmological parameters \citep{wmapest,wmap3}. On the other hand, this
quantity can be also determined combining data from  galaxy
surveys (SDSS, 2dF) and x-ray satellites (Chandra, XMM-Newton,
ROSAT, ASCA)  \citep{LMV06}. In this work, we consider a
weighted mean between the WMAP estimate and \citet{LMV06}
estimate for  $\Omega_b h^2$, and, furthermore, we shall compute
the dependence of binding energies, cross sections and abundances
on this parameter. Finally, we shall use observational data from
$\De$ and $\He$ to estimate the variations in time of $\alpha$ and
$<v>$ and a possible deviation of $\Omega_b h^2$ from its
considered value. We also  compare
our results with other non-SBBN models, where a non-standard expansion rate and an electron-neutrino
asymmetry were considered.
Finally, we would like to emphasize that 
the approach in this work is phenomenological and the results we get are
model independent.

The paper is organized as follows. In section 2 we present the
notations used and summarize the steps which one follows in the
semi-analytic approach to calculate the abundances. In section 3
we calculate the dependence of the abundances on $\alpha$, $<v>$,
$\Omega_b h^2$ and the deuterium binding energy $\epsilon_D$. In
section 4, we express the dependence of the deuterium binding
energy on the Higgs vev within the square well model. Results of
comparing theoretical prediction with observational values are
presented in section 5, where we also compare with other non-standard BBN models results. Conclusions are presented in section 6.

\section{Preliminaries}
\label{sec:Prel}

The method of \citet{Esma91} consists of calculating the different
abundances between fixed points or stages. One solves the
equations for the light elements
only for one element in each stage. For the other elements (say,
the $i^{th}$), it is necessary to solve the quasi static
equilibrium (QSE) equation ($\dot{Y_i} \approx 0$), where $Y_i$ is
the abundance of the $i^{th}$ element relatively to baryons,
considering only the most important rates of production and
destruction. On the other hand, we should also calculate the final
temperature of each stage. We show in Table \ref{resumen} the
different stages and their corresponding equations, to which is
added also the conservation of the neutron number (further details are given in \citet{Esma91, LMV06}).
\begin{table*}[h]
\caption{Stages and equations. $n$ refers to neutron, $p$ to
proton, $d$ to deuterium, $T$ to tritium, $3$ to $\Het$ and
$\alpha$ to $\He$ } \label{resumen}
\begin{tabular}{|c|c|c|}
\hline Stage& Equations &Final temperature \\ \hline

Until the weak interaction freeze-out&&\\ \hline

Until the production of \He~ becomes efficient&$
\dot{Y_n}=-2\dot{Y_{\alpha}}-Y_n[n] $&$2\dot{Y_{\alpha}}\sim
Y_n[n]$ \\ &$\dot{Y_d}=\dot{Y_3}=\dot{Y_T}=0$&  \\ \hline

Until the production of deuterium dominates&$\dot{Y_n}=-2\dot{Y_{\alpha}}$&$Y_n=Y_d$\\ 

rate of change of neutrons  &$\dot{Y_d}=\dot{Y_3}=\dot{Y_T}=0$&
\\ \hline

 Deuterium final abundance&$\dot{Y_d}=-2\dot{Y_{\alpha}}$&$T_9
\rightarrow 0$\\ &$\dot{Y_n}=\dot{Y_3}=\dot{Y_T}=0$&  \\ \hline

\end{tabular}
\end{table*}


Since we are considering changes in the Higgs vev ($<v>$) and the
fine structure constant ($\alpha$), we need to find expressions
for the nucleon masses and binding energies in terms of these
quantities. For the P-N mass splitting we have

\begin{equation}
Q=\Delta
m=m_n-m_p=\Delta^{\alpha} m+\Delta^{\rho-w} m
\end{equation}
where
$\Delta^{\alpha} m$ is the contribution of the electromagnetic
energy, and thus  $\frac{\delta \Delta^{\alpha}
m}{\Delta^{\alpha} m} = \frac{\delta \alpha}{\alpha}$. On the other hand,
$\Delta^{\rho-w} m$ is due to $\rho$-$w$ mesons mixing and known
to be proportional to $\frac{m_s^2}{m_u+m_d}$
\citep{Epele91b,Christiansen91}. Therefore,  $\frac{\delta
\Delta^{\rho-w} m}{\Delta^{\rho-w} m} = \frac{\delta m^q}{m^q}=
\frac{\delta <v>}{<v>}$. Thus, we get:
\begin{eqnarray}\label{DeltaQ}
\frac{\delta Q}{Q}=-0.587 \frac {\delta \alpha}{\alpha}+
1.587\frac{\delta <v>}{<v>}\end{eqnarray} We need also to know the
dependence of the variation of the nuclear mass of an element
$^A_ZX$ in terms of the changes in $<v>$ and $\alpha$. This can be
estimated using $M(X)=Zm_p+Nm_n-\epsilon_X$ where $\epsilon_X$ is
the binding energy for the element $X$, and we have:
\begin{equation} \label{energia} \frac{\delta
\epsilon_x}{\epsilon_x}=\frac{\epsilon_C}{\epsilon_x} \frac{\delta
\alpha}{\alpha}
\end{equation}
where $\epsilon_C=\frac{Z}{4 \pi \epsilon_0} \frac{e^2}{R}$ is the
electromagnetic contribution.
The radius of the nucleus ($R \sim 1.2 A^{\frac{1}{3}}fm$) is
considered as a strong interaction effect and, thus, taken to be
constant in our analysis. The change in the neutron decay rate in
terms of the changes in $\alpha$ and $<v>$ can be expressed as follows \citep{Ichi02,LMV06}:
\begin{eqnarray}
\label{DeltaTau}
\frac{\delta \tau}{\tau}=-3.838\frac {\delta
\alpha}{\alpha}-4.793\frac{\delta <v>}{<v>}
\end{eqnarray}
where we have used $\frac{\delta G_F}{G_F}=-2\frac{\delta
<v>}{<v>}$. For the thermonuclear reaction rates dependence on
$\alpha$, we take the phenomenological expressions of tables IV
and V in \citet{LMV06}.

Since BBN is very sensitive to $\epsilon_D$, we should go further
than equation (\ref{energia}) to evaluate the changes in
$\epsilon_D$ in terms of $\delta <v>$. We will give our
expressions for the different stages in terms of $\frac{\delta
<v>}{<v>}$,$\frac{\delta \epsilon_D}{\epsilon_D}$,$\frac{\delta
 \alpha}{\alpha}$ and $\frac{\delta \Omega_Bh^2}{\Omega_Bh^2}$. In section 4 we find an estimate for $\frac{\delta
 \epsilon_D}{\epsilon_D}$ in terms of $\frac{\delta
<v>}{<v>}$, and thus we can give then the final expressions in
terms of $\frac{\delta <v>}{<v>}$,  $\frac{\delta
 \alpha}{\alpha}$ and $\frac{\delta \Omega_Bh^2}{\Omega_Bh^2}$.

\section{Abundances and their dependence on $\alpha$ and $<v>$ in the different stages}



The ratio $X$ of the number of neutrons to the total number of
baryons in the first stage until the freeze-out of weak
  interactions ( $T > 9.1 \times 10^9 \rm K$) can be expressed as follows \citep{BBF88}:
\begin{eqnarray}
\label{xin}
X(y=\infty)=\int^\infty_0dy^{'}
e^{y^{'}}\frac{1}{1+e^y}\left(y^{'}\right)^2 e^{-K\left(y^{'}\right)}=0.151
\end{eqnarray}
where $K(y)=b\left[\frac{4}{y^3}+\frac{3}{y^2}+\frac{1}{y}+
\left(\frac{4}{y^3}+\frac{1}{y^2}\right)e^{-y}\right]$. Only
$b=255 \frac{M_{pl}}{\Delta m^2 \tau} \sqrt{\frac{45}{43\pi^3}}$
depends on the fundamental constants through $\tau$ and $\Delta
m$, so we get:
\begin{eqnarray}
\frac{\delta X(y=\infty)}{X(y=\infty)}=-0.52
\frac{\delta b}{b}
\end{eqnarray}
Using equations \ref{DeltaQ} and \ref{DeltaTau} we obtain:
\begin{eqnarray}
\label{deltayn1}
\frac{\delta
X(y=\infty)}{X(y=\infty)}&=&1.385\frac{\delta
\alpha}{\alpha}-0.842\frac{\delta <v>}{<v>}
\end{eqnarray}


In the second stage, after weak interactions freeze out, neutrons
decay freely until the rate of production of $^4He$ becomes
efficient ($9.1 \times 10^9 \rm K > T > 0.93 \times 10^9 \rm K $).
Thus we have:
\begin{eqnarray}
Y_n=X(y=\infty)\hskip
0.2cm e^{-t/\tau} =  \hskip 0.2cm 0.151 e^{-0.2/T_9^2}
\end{eqnarray}
where $T_9$ is the temperature evaluated in units of $10^9 K$.


The abundance of deuterium follows its equilibrium value and we
assume the reactions $[npd\gamma]$ and $[d\gamma np]$ dominate for
its production and destruction.
Taking $\Omega_B h^2 = 0.0223$ we can calculate the final
temperature of this stage by setting $\dot{Y_n}=0$ and thus
$2\dot{Y_n}=-Y_n[n]$. We find $T^f_9=0.93$ and get the abundances
$Y_p = 0.76$ and $Y_n = 0.12$.
In order to calculate the dependence of the final temperature on
the fundamental constants, we derive the equation
$2\dot{Y_n}=-Y_n[n]$ with respect to the fundamental constants to
get:
\begin{eqnarray}
\frac{\delta T_9}{T_9}&=&0.065\frac{\delta \Omega_B
h^2}{\Omega_Bh^2}+ 0.055 \frac{\delta \alpha}{\alpha}-0.119
\frac{\delta <v>}{<v>} + 1.195\frac{\delta \epsilon_D}{\epsilon_D} \nonumber
\end{eqnarray}
and thus we get the relative variations of the nucleons abundances
for this stage as follows:
\begin{eqnarray}
\frac{\delta Y_n}{Y_n}&=&0.030\frac{\delta \Omega_B h^2}{\Omega_Bh^2}+
2.159 \frac{\delta \alpha}{\alpha}-2.005
\frac{\delta <v>}{<v>}  + 0.553\frac{\delta \epsilon_D}{\epsilon_D} \nonumber\\
\label{ypfinal} \frac{\delta Y_p}{Y_p}&=&
-0.009\frac{\delta\Omega_B h^2}{\Omega_B h^2}
-0.682 \frac{\delta
\alpha}{\alpha}+0.634 \frac{\delta <v>}{<v>} - 0.174\frac{\delta
\epsilon_D}{\epsilon_D}\nonumber
\end{eqnarray}

In order to compute the final abundance of helium, we notice that
once $^4He$ production becomes efficient (i.e.
$2\dot{Y_{\alpha}}=Y_n [n]$), neutrons combine to form
$\alpha$-particles, and the production of the latter is dominated
by $[dTn\alpha]$ and $[pT\gamma \alpha]$.
One gets for the temperature of the $^4He$ freeze-out the
following equation:
\begin{eqnarray}
\label{finalalpha} 2 Y_n \left(Y_p \frac{[npd\gamma]}{Y_{\gamma}[d
\gamma np]}\right)^2 [ddpT] =\frac{1}{\tau}
\end{eqnarray}
where $Y_p=0.76$, $Y_n=0.151 e^{-0.2/T_9^2}$ and $\tau$ is the
neutron decay constant. Numerically we find $T_9^{\alpha}=0.915$
which is lower than the final temperature of the previous stage
and larger than the final temperature of the next one. For the
final helium abundance we find $Y_{\alpha}^f=2Y_n=0.238$. As
before, deriving equation \ref{finalalpha} with respect to
$\epsilon_D$, $<v>$ and $\alpha$ we find:
\begin{eqnarray}
\frac{\delta T_9^{\alpha}}{T_9^{\alpha}}&=&0.061\frac{\delta
\Omega_B h^2}{\Omega_B h^2}+ 0.049 \frac{\delta
\alpha}{\alpha}-0.113 \frac{\delta <v>}{<v>} + 1.149\frac{\delta
\epsilon_D}{\epsilon_D}\nonumber
\end{eqnarray}
Since $Y_{\alpha}^f=2 Y_n$ we get the relative variation of the
helium abundance as:
\begin{eqnarray}
\frac{\delta Y_{\alpha}^f}{Y_{\alpha}^f}&=&0.029\frac{\delta\Omega_B
h^2}{\Omega_B h^2}+ 2.182
\frac{\delta\alpha}{\alpha}-2.042\frac{\delta
<v>}{<v>} + 0.549\frac{\delta \epsilon_D}{\epsilon_D} \nonumber
\end{eqnarray}



In the following `neutron cooking' stage, corresponding to $0.93
\times 10^9 \rm K
> T
> 0.766 \times 10^9 \rm K$, the neutron abundance can be expressed
as follows:
\begin{eqnarray} \label{integral}
Y_n=\left(\frac{1}{Y_n^0}+2\int^{t}_{t_{initial}} \left(Y_p
\frac{[npd\gamma]}{Y_{\gamma}[d\gamma np]}\right)^2 [ddpT]
dt\right)^{-1}
\end{eqnarray}
where the initial condition is given by the final values of the
previous stage: $Y_n^0=0.12$ at $T_9^0=0.93$.


Putting $Y_n=Y_d$ as the condition which determines the final
temperature of this stage, we find
$\frac{[npd\gamma]}{Y_{\gamma}[d\gamma np]} = 1$.
With $Y_p$ freezed at 0.76, we get numerically:
\begin{eqnarray}
T_9^f&=&0.766\\
Y_n&=&6.4\times 10^{-4}=Y_d
\end{eqnarray}
Again, the condition ($Y_n=Y_d$) allows the calculation of the
relative change of the final temperature:
\begin{eqnarray}
\frac{\delta T_9^f}{T_9^f}&=&0.031 \frac{\delta\Omega_B h^2}{\Omega_B h^2}
-0.021 \frac{\delta \alpha}{\alpha}+0.020\frac{\delta <v>}{<v>} +1.041\frac{\delta
\epsilon_D}{\epsilon_D}\nonumber
\end{eqnarray}
and we get numerically:
 \begin{eqnarray}
\frac{\delta Y_d}{Y_d}=\frac{\delta Y_n}{Y_n}&=&-1.095 \frac{\delta
\Omega_Bh^2}{\Omega_B h^2} + 1.865 \frac{\delta
\alpha}{\alpha}-0.075\frac{\delta <v>}{<v>} - 2.275\frac{\delta
\epsilon_D}{\epsilon_D}\nonumber
\end{eqnarray}


In the last stage ($T < 0.766 \times 10^9 \rm K$), we notice that
the dominant term in the time derivative of $Y_d$ is the
production of Tritium, i.e. $Y_d Y_d [ddpT]$.
We find:
 \begin{eqnarray}
\label{ydfinal}
 Y_d=\left(\frac{1}{Y_d^0}+2\int^{t}_{t_{initial}}
[ddpT] dt\right)^{-1}
\end{eqnarray}
with the initial value $Y_d^0=6.4 \times 10^{-4}$ at
$T_9^0=0.766$.

We can obtain the final abundance of deuterium by setting the
temperature equal to zero and we find the abundance numerically
equal to $Y_d^f=2.41 \times 10^{-5}$. Again, we can numerically
evaluate the relative change of $Y_d^f$:
\begin{eqnarray}
\label{variacion de yd final} \frac{\delta
Y_d^f}{Y_d^f}&=&-1.072\frac{\delta \Omega_B h^2}{\Omega_B h^2}+
2.318 \frac{\delta \alpha}{\alpha}-0.049\frac{\delta
<v>}{<v>}- 2.469\frac{\delta \epsilon_D}{\epsilon_D} \nonumber
\end{eqnarray}
We summarize the results that we obtained in table \ref{varia}.


\begin{table}[h]
\caption{Final abundances and their relative variations in terms
of the relative variations of the fundamental constants,
$\frac{\delta Y_i^f}{Y_i^f}=A \frac{\Omega_B h^2}{\Omega_B
h^2}+B\frac{\delta \alpha} {\alpha}+ C\frac{\delta <v>}{<v>}+ D
\frac{\delta \epsilon_D}{\epsilon_D}$} \label{varia}
\begin{center}
\begin{tabular}{|c|c|c|c|c|c|}
\hline $Y_i^f$&$A$&$ B$&$C$&$D$&Abundance\\ \hline

$D $&$-1.072$&$2.318$&$-0.049$&$-2.469$&$2.41 \times 10^{-5}$  \\
\hline

$^4He $&$0.029$&$2.182$&$-2.042$&$0.549$&$0.238$ \\ \hline
\end{tabular}
\end{center}
\end{table}


\section{The dependence of deuterium  binding energy $\epsilon_D$ on the Higgs vev $<v>$}
\label{dbe}
As we said before, the deuterium  binding energy $\epsilon_D$ is
the most significant factor that can influence the BBN reactions
rates, and its variation was discussed in
\citet{Flambaum02,Flambaum03,Dmitriev03,Flambaum04}. Indeed, the
equilibrium concentration of deuterons and the inverse reaction
rate depend exponentially on it. Moreover, the deuterium  is a
shallow bound level ($\epsilon_D \approx 2.225 \rm MeV$).
Therefore the relative variation of the deuterium binding energy
$\epsilon_D$ is much larger than the relative variation of the
strong interaction potential which we neglect in our work. In
order to give an estimate for the relative variation of
$\epsilon_D$, we should compute, within  perturbation
theory, the correction to $\epsilon_D$ due to the perturbation
which might change in time. Thus we write
$\epsilon_D=\epsilon_D^0+\Delta E$, where $\epsilon_D^0$ is the
unperturbed binding energy and we consider it a QCD-determined
quantity which does not change in time. As to $\Delta E$, we know
\citep{Weinberg90,Weinberg91} that the one-pion exchange potential
represents the first approximation to the perturbation on the
strong interaction potential, and it has the form:
\begin{eqnarray}
\label{Yukawa}
V^Y &=& \frac{f^2}{4\pi}\frac{e^{-m_\pi r}}{r}
\end{eqnarray}
where $\frac{f^2}{4\pi}  \sim 0.08$, $m_\pi\sim 140 \rm MeV$ is the pion mass.
We simplify the strong interaction potential by a square
well model with width $a$ and depth $V_0$. These two parameters
can be determined by fitting the square well `theoretical'
expressions involving these two parameters to the p-n scattering
data. According to the shape-independent effective range theory
\citep{Bethe49,Bethe50} all the binding and low energy scattering
properties of the potential are determined by just two parameters
which can be determined experimentally: the scattering length $a_t
= 5.50 \times 10^{-13} \rm cm$ and the effective range  $r_t =1.72
\times 10^{-13} \rm cm$ \citep{Schiff68}.
The corresponding
values for the square well are: the depth $V_0=35.5 \rm {MeV}$ and
the width $a=2.03 \times 10^{-13} \rm{cm}=0.0103 \rm{MeV}^{-1}$.
We consider the width and the depth as QCD-determined parameters
and assume they do not change in time.

Now, we have in the square well model
\begin{eqnarray}
\Delta E&=&\frac{f^2 A^2}{4 \pi \xi^2}\int_0^{\xi a} \frac{\sin^2
x}{x} e^{-\frac{m_\pi}{\xi}x} dx \nonumber \\ &&+\frac{ f^2 B^2 }{4 \pi
   \beta^2}\int_{\xi a}^{\infty}
\frac{e^{-\left(\frac{m_{\pi}}{\xi}+\frac{2\beta}{\xi}\right)x}}{x}
dx
\end{eqnarray}
where:
\begin{eqnarray}
\xi&=&\sqrt{m_N (V_0- \epsilon_D)}=176.76  \hspace{0.2cm}\rm{MeV}\\
\beta&=&\sqrt{m_N \epsilon_D }=45.71  \hspace{0.2cm}\rm{MeV}\\
A&=&\sqrt{\frac{\xi^2}{\frac{a}{2}-\frac{sin(2\xi a)}{4
\xi}+\frac{sin^2(\xi a)}{2 \beta}}}=1393 \hspace{0.2cm} \rm{MeV}^{3/2}\\
B&=&-\frac{\beta}{\xi} \sin{(\xi a )}e^{\beta a
}A=-559  \hspace{0.2cm}\rm{MeV}^{3/2}
\end{eqnarray}

and $m_N$ is the reduced mass of the two-nucleon system.

Whence,
\begin{eqnarray}
\Delta E=0.203\rm \hspace{0.2cm}{MeV}
\end{eqnarray}

On the other hand, the change due to the variation of the nucleon
mass in the Yukawa potential $V^Y$ is negligible compared with the
change due to the variation of $m_\pi$. Since $\frac{\delta
m_\pi}{m_\pi}=\frac{1}{2} \frac{\delta <v>}{<v>}$ as mentioned
before, we get, evaluating numerically the integrals, the
following:
\begin{eqnarray}
\frac{\delta \Delta E}{\Delta E}&=&-0.896\frac{\delta
m_\pi}{m_\pi}=-0.448\frac{\delta <v>}{<v>}
\end{eqnarray}
and so,
\begin{eqnarray}
\frac{\delta \epsilon_D}{\epsilon_D}&=& -0.041 \frac{\delta
<v>}{<v>}
\end{eqnarray}
These values are one order of magnitude lower than those obtained
by \citet{Flambaum02}. The difference arises from the fact that
these authors did not consider the continuity of the wave function
and its derivative on the boundary of the square well. This
results in differences between the normalization factor of the
wave functions which propagate into the binding energy first order
perturbation. Moreover, the values of $a$ and $V_0$ they consider are different from this work.

Hence, the final expressions for the relative variations of the
helium and deuterium abundances are:
\begin{eqnarray} \label{FinalRes1}
 \frac{\delta Y_d^f}{Y_d^f}&=&-1.072\frac{\delta \Omega_B
 h^2}{\Omega_B h^2}+2.318 \frac{\delta
 \alpha}{\alpha}+0.052\frac{\delta <v>}{<v>} \\
\label{FinalRes2} \frac{\delta
Y_{\alpha}^f}{Y_{\alpha}^f}&=&0.029\frac{\delta \Omega_B
h^2}{\Omega_B h^2}+2.182 \frac{\delta
\alpha}{\alpha}-2.044\frac{\delta <v>}{<v>}
\end{eqnarray}
These results are summarized in table \ref{fvaria}

 \begin{table}[h]
 \caption{Abundances and their dependence on fundamental constants,
 $\frac{\delta Y_i^f}{Y_i^f}=A\frac{\delta \alpha} {\alpha}+B
 \frac{\delta <v>}{<v>}+C\frac{\delta \Omega_B h^2}{\Omega_B h^2}$}
 \label{fvaria}
 \begin{center}
 \begin{tabular}{|c|c|c|c|c|}
 \hline $Y_i^f$&Abundance&$A$&$B$&$C$\\ \hline

 $\D $&$2.41 \times 10 ^{-5}$&$2.318 $&$0.052$&$-1.072$\\
 \hline

 $^4He$&$0.238$&$2.182 $&$-2.044 $&$0.029$\\ \hline
 \end{tabular}
 \end{center}
 \end{table}

\section{Results}

We can now compare the theoretical predictions of the abundances
of $^4He$ and $D$ obtained in the last section with the
observational data. The equations (\ref{FinalRes1},
\ref{FinalRes2}) are of the form ($i=1(D),2(^4He)$)
\begin{eqnarray}
\label{FitEq} \frac{\delta Y_i^f}{Y_i^f}&=&A_i\frac{\delta
\Omega_B h^2}{\Omega_B h^2}+B_i\frac{\delta \alpha} {\alpha}
+C_i\frac{\delta <v>}{<v>}
\end{eqnarray}
and we take the assumption that the difference $\frac{\delta
Y_i^f}{Y_i^f}$ is due to a change in the considered fundamental
constants: $\frac{\delta
Y_i}{Y_i}=\frac{Y_i^{obs}-Y_i^{SBBN}}{Y_i^{SBBN}}$, where
$Y_i^{SBBN}$ and $Y_i^{obs}$ are the theoretical and observed
abundances respectively. In table \ref{SBBN}, the theoretical abundances $Y_i^{SBBN}$ are
given for $\Omega_B h^2=0.0223$ with their errors resulting from
the uncertainty in the values of the parameters involved.
In table \ref{obs}, the observational data of helium and deuterium
are stated with their measured errors.


\begin{table}[h]
\caption{Theoretical
abundances in the standard model with the WMAP estimate  $\Omega_B
h^2=0.0223$} \label{SBBN}
\begin{center}
\begin{tabular}{|c|c|}
\hline Nucleus & $Y_i^{SBBN} \pm \delta Y_i^{SBBN}$  \\ \hline
$\D$&$\left(2.51\pm 0.37\right)\times 10^{-5}$\\ \hline
$\He$&$0.2483\pm 0.0021$\\\hline
\end{tabular}
\end{center}
\end{table}

\begin{table}[h]
\caption{Observational abundances} \label{obs}
\begin{center}
\begin{tabular}{|c|c|c|}
\hline

Nucleus & $Y_i^{obs} \pm \delta Y_i^{obs}$ & Reference  \\ \hline

D& $\left (1.65 \pm 0.35\right ) \times 10^{-5}$&\cite{pettini}\\
\hline

D&$\left (2.54 \pm 0.23\right ) \times 10^{-5}$&\cite{omeara}
\\ \hline

D&$\left (2.42^{+0.35}_{-0.25}\right ) \times
10^{-5}$&\cite{kirkman} \\ \hline

D&$\left (3.25 \pm 0.3\right ) \times 10^{-5}$& \cite{burles2}\\
\hline

D&$\left (3.98^{+0.59}_{-0.67}\right ) \times 10^{-5}$&
\cite{burles1}\\ \hline

D&$\left (1.6^{+0.25}_{-0.30}\right ) \times 10^{-5}$&
\cite{Crighton04}\\ \hline

$^4\rm{He}$& $0.244 \pm 0.002$&\cite{izotov1} \\ \hline

$^4\rm{He}$&$0.243 \pm 0.003$&\cite{izotov2}\\ \hline

$^4\rm{He}$&$0.2345 \pm 0.0026$ &\cite{peimbert1} \\ \hline

$^4\rm{He}$&$0.232 \pm 0.003$&\cite{oli95}\\ \hline

\end{tabular}
\end{center}
\end{table}


As regards the consistency of the $\De$ and $^4He$ data, we follow the treatment of \citet{LMV06} and
increase the observational error by a factor
$\Theta$. The values of $\Theta$ are $2.4$ for $\De$, $2.33$ for
$\He$.

The results of solving the system of equations (\ref{FitEq}) with
the given data are shown in table \ref{square}. These results are
consistent within $1-\sigma$ with no variation of the fundamental
constants. On the other hand, the results considering variation of
one fundamental constant only are shown in table \ref{square2}.
These results are consistent within $3-\sigma$ with no variation
of the fundamental constants.

 \begin{table}[h]
 \caption{Results of the analysis using the square well model considering
 joint variation of all constants}
 \label{square}
 \begin{center}
 \begin{tabular}{|c|c|c|}
 \hline
 Realtive variation & Value& $\sigma$\\ \hline

 $\frac{\delta \Omega_B h^2}{\Omega_B h^2}$&-0.011&0.054\\ \hline

 $\frac{\delta \alpha}{\alpha}$&-0.032 &0.072\\
 \hline

 $\frac{\delta <v>}{<v>}$&-0.011 &0.078\\ \hline
 \end{tabular}
 \end{center}
 \end{table}

 \begin{table}[h]
 \caption{Results of the analysis using the square well model considering variation
 of each constant only}
 \label{square2}
 \begin{center}
 \begin{tabular}{|c|c|c|}
 \hline
 Realtive variation & Value& $\sigma$\\ \hline

 $\frac{\delta \Omega_B h^2}{\Omega_B h^2}$&-0.013&0.056\\ \hline

 $\frac{\delta \alpha}{\alpha}$&-0.015 &0.006\\
 \hline

 $\frac{\delta <v>}{<v>}$&0.017 &0.007\\ \hline
 \end{tabular}
 \end{center}
 \end{table}





\begin{figure}[h!]
\begin{center}
\label{residuos} \caption{The full line shows the theoretical
probability of the residuals, while the dotted line shows the
empirical probability. Deviations of $\Omega_B h^2$, $\alpha$ and
$<v>$ with respect to their mean values are considered}
\epsfig{file=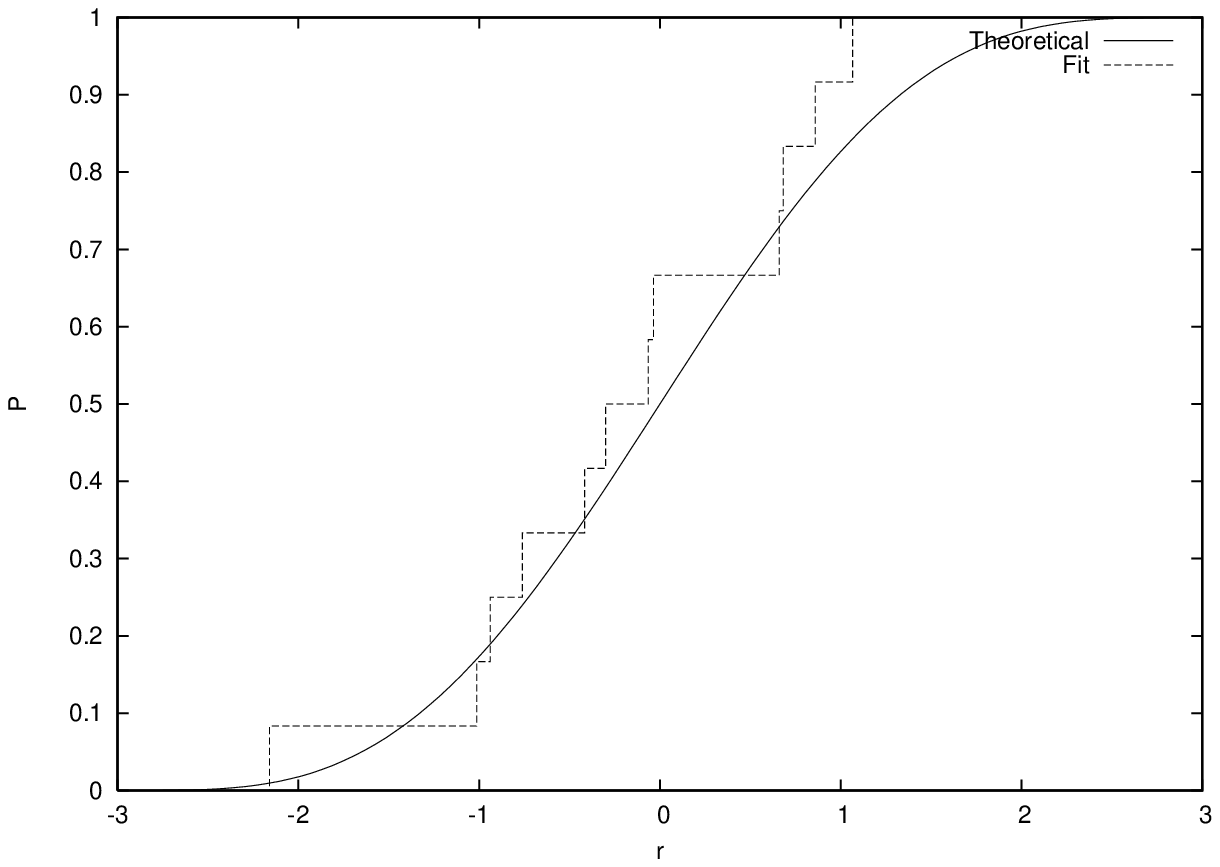, angle=0, width=200pt}
\end{center}
\end{figure}

In order to check the goodness of our fit, we performed a
Kolmogorov-Smirnov (K-S) test (see figure \ref{residuos}). When
considering  variation in $\Omega_B h^2$, $\alpha$ and $<v>$
altogether, we have a probability of $79\%$ to obtain a worse fit.
However, we consider the results of the K-S test only indicative,
since even though the considered data are independent the
residuals are not.

Finally, it is interesting to compare our results with other non-standard BBN models. In particular, a non-standard expansion rate and  an electron-neutrino asymmetry were considered by several authors \citep{Barger03,Steigman05,Steigman06}. While $\De$ is more sensitive to the baryon density ($\Omega_b h^2$), the effect of a non zero $\nu_e$ asymmetry is more strong for $\He$ than for the other relic nuclides. This is similar to the effect of $< v >$ as follows from equations \ref{FinalRes1} and \ref{FinalRes2}, whereas both $\De$ and $\He$ are sensible to changes in $\alpha$. In the papers cited above, the authors considered the following cases: i) adding only one free parameter (either $\Delta N_{\nu}$ or $\xi_e$)  and ii) adding two free parameters ($\Delta N_{\nu}$ and $\xi_e$) besides the baryon density. They consider the WMAP data from the CMBR togheter with $\He$ and $\De$ abundances. In the first case, the results are consistent within $1-\sigma$ with  $\xi_e \neq 0$ and $\Delta N_{\nu} \neq 0 $. In the second case, the results are consistent within $1-\sigma$ with $\xi_e = 0$ and $\Delta N_{\nu}=0$. These results are similar to ours in that considering only one free parameter besides the baryon density is consistent with non standard physics within $1-\sigma$ whereas considering two free parameters is consistent within $1-\sigma$ with the SBBN model. However, it is important to remind that most of the theories where the fundamental constants may vary in cosmological time scales, predict joint variation of constants.

\section{Conclusions}
\label{conclusion} In this work, we assumed that the discrepancy
between SBBN estimation for $^4He$ and $D$ and their observational
data is due to a change in time for the fundamental constants: the
Higgs vev $<v>$, the fine structure constant $\alpha$. We analyzed the dependence of the
$^4He$ and $D$ abundances on these fundamental constants within
 perturbation theory and on deviations with respect to  the mean value of the baryonic density. Furthermore, we  compared them with the
observational data. We find that varying fundamental constants may
not solve, in our case, the discrepancy between the theoretical
SBBN and the observed data considered in this work. We hope this work stimulates further research in this interesting subject.

\section*{{\bf Acknowledgements}}
Support for this work is provided by Project G11/G071, UNLP.
H.Vucetich is partially sponsored by Project 42026-F of CONACyT, Mexico.
N.Chamoun wishes to acknowledge support from TWAS and CONICET, Argentina


\bibliography{bibliografia2}
\bibliographystyle{astron}

\end{document}